\documentclass[referee]{jfm2}
\usepackage{amsmath, amssymb}
\usepackage[dvips]{graphicx}
\usepackage{natbib}
\usepackage{bm}

\def\C{{\mathcal C}}

\def\hf{{\frac{1}{2}}}

\newtheorem{thm}{Theorem}[section]

\newtheorem{defn}[thm]{Definition}

\begin{document}

\bibliographystyle{plainnat}
\title[Trapped slender vortex filaments]{Trapped slender vortex filaments in statistical equilibrium}
\author[T. D. Andersen and C. C. Lim]{Timothy D. Andersen \thanks{andert@rpi.edu}
 \and Chjan C. Lim \thanks{limc@rpi.edu}}

\affiliation{Mathematical Sciences, RPI, 110 8th St., Troy, NY, 12180}

\date{\today}

\begin{abstract}
The statistical mechanics of nearly parallel vortex filaments confined in the unbounded plane by angular momentum, first studied by Lions and Majda (2000), is investigated using a mean-field approximation to interaction and a spherical constraint to develop an explicit formula for the mean square vortex position or length scale of the system, $R$, verified with
Path Integral Monte Carlo simulations.  We
confirm that 3D filaments resist confinement in a different way than 2D point vortices and that this results in a
profound shift at high-densities for the length scale of quasi-2D versus strictly-2D models of vorticity fields in which angular momentum is conserved.  Our analytical
results correspond well with those of the Monte Carlo simulations and show a 3D effects
contributing significantly to determination of the length scale.
\end{abstract}

\maketitle

\section{Introduction}
Statistical mechanics is a way of calculating the macroscopic properties of matter from the
probabilistic behavior of its microscopic components.  Rather than solving the Navier-Stokes equations
explicitly in time, a statistical equilibrium approach for fluid flows aims to describe observable quantities by
averaging over ``microstates'' or states that account for a system's exact position in
phase space (\cite{Majda:2006}).

In fluid turbulence a statistical treatment is often preferable
to obtaining direct solutions to the Navier-Stokes equations because of the inherent chaos/complexity of turbulent
flows that makes it impossible to model every trajectory at high Reynolds numbers.  
This nonequilibrium statistical description of turbulence can be replaced by a statistical equilibrium approach under specific conditions such as nearly inviscid quasi-2D flows where a separation of time scales is valid. Familiar arguments in favor of a separation in the time scales of energy and angular momentum transfer at flow boundaries versus viscous dissipation are based on the tea cup paradigm.
The ``universal equilibrium assumption'', that the time scale at which turbulent features form is much
smaller than the time scale for viscous decay, justifies studying turbulent flows in equilibrium (\cite{Chorin:1994}). Although this justification can be considered tenuous and subject to mathematical
limits, experimental observations provide a validation for the statistical approach.

Most work on fluid flows in statistical equilibrium is on how large scale structures appear in vorticity fields of 2D rotating ideal fluids such as Onsager's Point Vortex Gas (\cite{Onsager:1949}).  Many significant results have come out of this approach, but research has confirmed few 2D results for nearly-2D or quasi-2D models.  In many cases,
2D models approximate a quasi-2D reality.  This paper is concerned with when this approximation fails
and how the statistics of quasi-2D models depart from those of fully-2D when it does fail.

\begin{figure}
\begin{center}
\includegraphics[width = \textwidth]{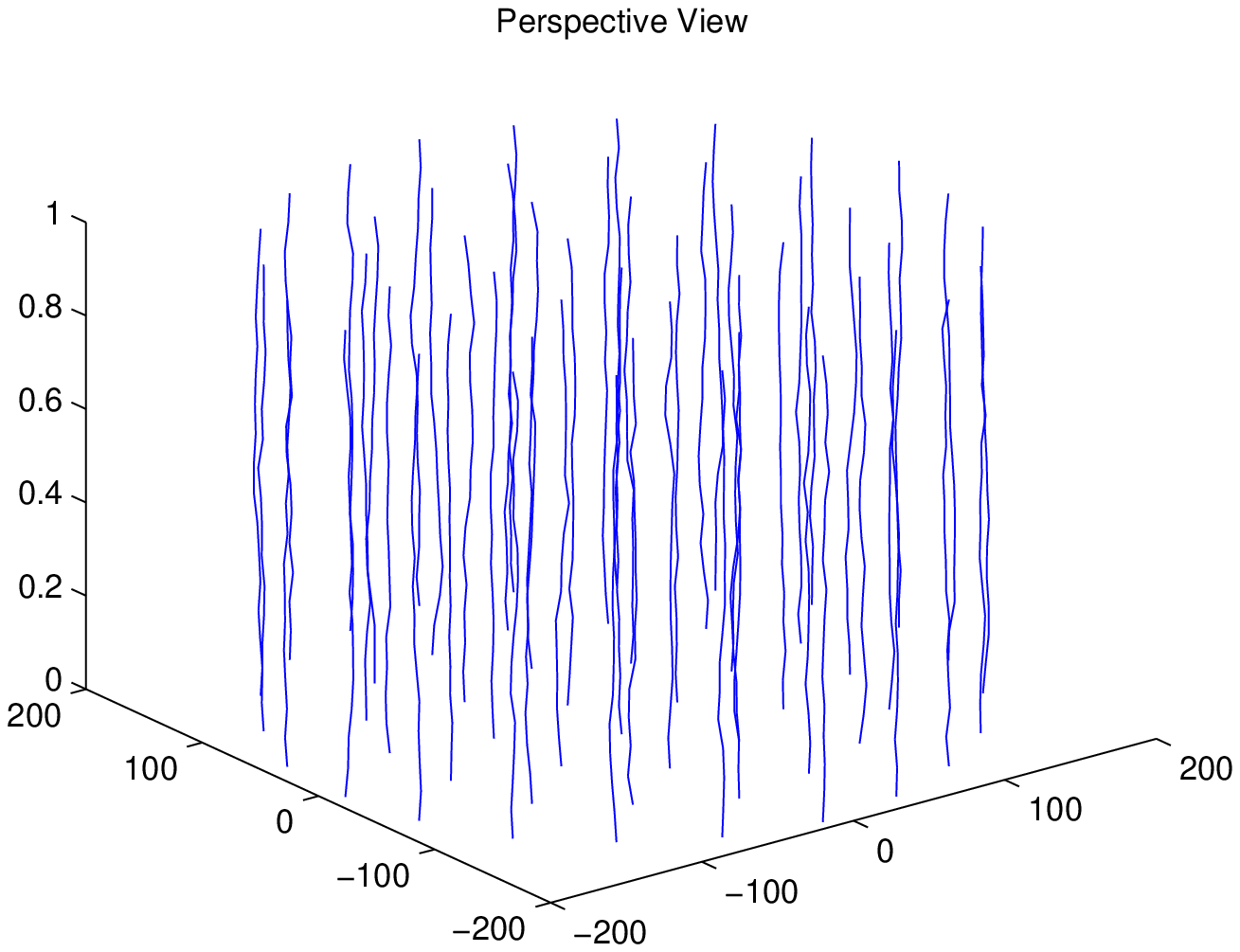}
\end{center}
\caption{Nearly parallel vortex filaments are nearly parallel to the z-axis in an asymptotic sense
that their deviation from straight is a small parameter.  They are infinite in length but have a
period $L$.  Here there are 50 filaments.}
\label{fig:paths3d}
\end{figure}

Nearly parallel vortex filaments (Figure \ref{fig:paths3d}) are one of the simplest model for discrete, quasi-2D vorticity. 
They appear in an incompressible, nearly inviscid Navier-Stokes flow (\cite{Klein:1995}) as well as other physical systems such as in rotating superfluids and Bose-Einstein condensates.  From theories for single vortex filaments by \cite{Hasimoto:1972},\cite{Callegari:1978}, \cite{Ting:1991}, and \cite{Klein:1991}, \cite{Klein:1995} developed the first rigorous model for \emph{interacting} nearly parallel filaments.  \cite{Julien:1996} have observed these nearly parallel vortex filaments in astrophysical simuations.

\begin{figure}
\begin{center}
\includegraphics[width = \textwidth]{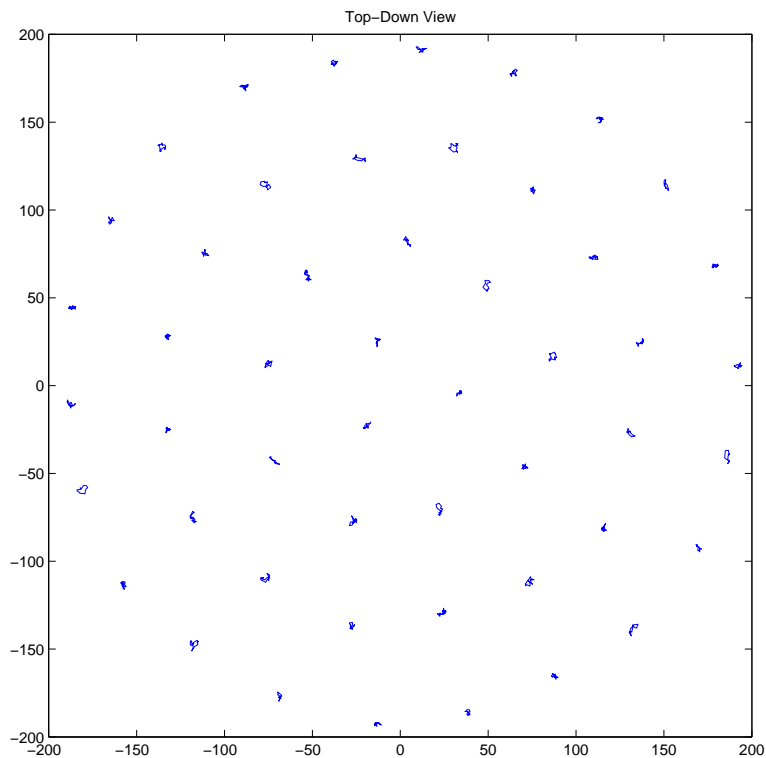}
\end{center}
\caption{Shown here in top-down projection, nearly parallel vortex filaments, at low density and high strength of interaction,
are well-ordered into a 2-D triangular lattice known as the Abrikosov lattice from type-II superconductors (\cite{Abrikosov:1957}).  This figure shows how the quasi-2D model is essentially
a 2-D model for these parameters.}
\label{fig:paths2d}
\end{figure}

Our statistical description of nearly parallel vortex filaments derives from the work of \cite{Lions:2000}.  In their paper, they developed a description of the time-dependent statistics, namely the density, $\rho(t,{\bf r})$, where ${\bf r}\in \Re^2$,
of the filaments for $N\rightarrow\infty$, given that energy and angular momentum are conserved, using the model of \cite{Klein:1995}.  Filaments are also periodic in the z-direction with period $L$.  

They introduce a Gibbs probability measure for the vortex filaments in the unbounded plane, confined by angular momentum.  
This distribution models a system that is not completely isolated, a reasonable assumption in geophysical and other macroscopic flows, in that a quiescent fluid surrounding the filaments acts as an energy and angular momentum bath.  The system is allowed to exchange angular momentum and energy with the environment and so angular momentum and energy fluctuate. A justification, \emph{a posteriori}, of the Gibbs' probability measure that is canonical, rather than microcanonical, in energy and angular momentum comes from results indicating that
vortices are confined to a compact domain in the plane, and the heat bath exists outside
this domain.
In formal notation, a canonical Gibbs' measure means that if a state $s$ has energy $E_s$ and angular momentum $I_s$, then the probability of $s$,
\begin{equation}
 P(E_s,I_s) \propto e^{-\beta E_s - \mu I_s},
\label{eqn:gibbs}
\end{equation} where $\beta$ and $\mu$ are constants.  (For historical reasons $\beta$ is known as inverse ``temperature'' and $\mu$ as ``chemical potential'' although they are not directly related to their molecular equivalents because the ``molecules'' here are vortices.)  Since $\sum_s P(E_s, I_s) = 1, P(E_s,I_s) = Z^{-1}e^{-\beta E_s - \mu I_s}$, where $Z = \sum_s e^{-\beta E_s - \mu I_s}$ (\cite{Chorin:1994}).  Lions' and Majda's
rigorous treatment of this statistical ensemble leads to a non-linear Schroedinger equation governing
$\rho$.  Their approach has been criticised as being physically inconsistent because it admits states that are
not nearly parallel (\cite{Berdichevsky:2002}).  This criticism is valid.  However, a careful choice of parameters $\beta$, $\mu$, etc. causes these disallowed states to have such large energies that their probabilities are negligible. For such parameter ranges, we can assume that the Lions-Majda model is physical.  Our Monte Carlo simulation naturally rejects states
with non-nearly-parallel lines without additions to the model.

The conservation of angular momentum is key in this study because it introduces a natural
length scale to the confined system of vortices in the unbounded plane (\cite{DiBattista:2001}).  It also plays an important role in statistical mechanics on the sphere (\cite{Majda:2006},\cite{Lim1:2006},\cite{Lim2:2006}).  Periodic boundary conditions are more common, but
these enforce a length scale (the period) artificially on the plane that can affect the statistics.
Although a pre-determined length scale is reasonable in some cases, for many others, it is not.
Moreover, the statistics for periodically bounded ensembles is quite different from unbounded systems that are confined by angular momentum or a harmonic trap as in the case of Bose-Einstein condensates. As the
reader will see in this paper, our results, which focus on determining the length scale both
analytically and computationally, depend upon the angular momentum confinement.

Our goal is not merely to demonstrate that nearly parallel vortex filaments behave differently from 2D point vortices.  Nearly parallel vortex filaments at low-density/high-straightness (large $\beta$ or low-temperature) behave \emph{exactly} like 2-D point vortices (Figure \ref{fig:paths2d}).  That their behavior at high densities and levels of curvature (small $\beta$ or high-temperature) ought to be different than that of 2D point vortices is obvious.  On the other hand, despite the rigorous work of Lions and Majda, the qualitative statistical differences between nearly parallel vortex filaments and strictly-parallel filaments or 2D point vortices are unknown.

Our hypothesis is that at high-densities point vortices and nearly parallel vortex filaments have qualitatively different behavior.  The length scale of point vortices collapses with increasing temperature (\cite{Lim:2005}).  We hypothesize that filaments reverse this collapse (Sec. \ref{sec:results}).  Similarily to how stars in globular clusters
resist gravitational collapse through motion, we posit that filaments resist collapse through curvature (which in top-down projection appears like Brownian motion) and that the outcome is a reversal of
collapse as temperature increases.  This idea has profound implications for the natural length scale
of 2D versus quasi-2D models.

We develop a simpler approach than
Lions and Majda---one that is motivated and justified 
by their rigorous mean-field result---that we hope generates a more intuitive understanding of what happens when the density
and curvature of the filaments increases and interaction becomes more three dimensional.  Our method
is two-fold:
\begin{enumerate}
 \item We develop an explicit formula for the second moment, $R^2$, (i.e. the square length scale) of the equilibrium vortex density distribution, $\rho(t\rightarrow\infty,{\bf r})$, using a simpler mean-field approximation to interaction than Lions-Majda.  Our mean-field approach assumes that the interaction due to $N-1$ filaments on one filament is similar to how a perfectly straight filament fixed at the origin with strength $N-1$ would affect the the center-of-mass of that filament.  This assumption requires that the filaments be fairly uniformly distributed, whereas Lions-Majda do not require uniformity.  It is a ``particle interacting with a center-of-mass'' rather than ``particle interacting with a density field'' assumption.  We then constrain the filament's planar position such that,
\begin{equation}
 \int_0^L d\tau |\Psi(\tau)|^2 = LR^2,
\end{equation} where $\Psi(\tau)=x(\tau) + iy(\tau)$ represents the curve of a filament in complex notation.  This constraint enters the Gibbs distribution (Eqn. \ref{eqn:gibbs}) as an exact (micro-canonical) conservation law.  These two assumptions, discussed in detail below, allow us to derive the formula:
\begin{equation}
 R^2 = \frac{\beta^2\alpha N + \sqrt{\beta^4\alpha^2N^2 + 32\alpha\beta\mu}}{8\alpha\beta\mu},
\label{eqn:rsq3d}
\end{equation} where $\alpha$ is related to core structure.  (Larger $\alpha$ means straighter filaments.)  (Section \ref{sec:meanfield}.)

\item To confirm our formula, we perform Path Integral Monte Carlo (\cite{Ceperley:1995}) on the original statistical system using piecewise linear approximations to the filaments.  These simulations also
allow us to determine that our results are physical, addressing the point of criticism by 
\cite{Berdichevsky:2002}.  (Section \ref{sec:montecarlo}.)
\end{enumerate}

Our statistical derivation not only answers the questions above but gives an extremely close prediction of length scale, $R$, and allows us to explain the qualitative differences between
a strictly-2D and a nearly-2D model, giving new insight into the role of 3D effects in the onset of turbulence in quasi-2D models.

The paper is laid out as follows: we discuss the problem in greater detail, including the 2D Onsager Gas (\S \ref{sec:problem}), go over the model presented in \cite{Lions:2000}
(\S \ref{sec:model}), present our mean-field derivation for $R^2$ (\S \ref{sec:meanfield}), a description of Path Integral Monte Carlo (\S \ref{sec:montecarlo}), and a
comparison of Monte Carlo results with the mean-field formula and a verification of
the assumptions of the model (\S \ref{sec:results}).  We discuss and conclude (\S \ref{sec:discussion}, \S \ref{sec:conc}).

\section{Problem}
\label{sec:problem}
The low-temperature (asymptotic as $\beta\rightarrow\infty$) statistics of strictly-2D point vortices in the unbounded plane with angular momentum conserved are well understood.  
The Point-Vortex Gas Model has the Hamiltonian,
\begin{equation}
 H_N^{2D} = -\sum_{j>i}\lambda_i\lambda_j\log|\Psi_i - \Psi_j|^2,
\end{equation} where $\lambda_i$ and $\lambda_j$ are the circulation constants for point vortices $i$ and $j$ and $\Psi_i$ and $\Psi_j$ are their positions in the complex plane.  This Hamiltonian derives from the 2D Euler equations in vorticity form given that $\omega$, the field, has the form
\[
 \omega(\Psi) = -\sum_{j}\lambda_j\delta(\Psi - \Psi_j),
\] where $\Psi$ is a position in the complex plane.
 (\cite{Chorin:1994}).  The probability distribution,
\begin{equation}
 P_N = \frac{\exp{(-\beta H_N - \mu I_N)}}{Z_N},
\end{equation} where
\begin{equation}
 Z_N = \int d\Psi_1\cdots\int d\Psi_N \exp{(-\beta H_N - \mu I_N)}
\label{eqn:partfn}
\end{equation} and 
\begin{equation}
 I_N = \sum_i^N \lambda_i|\Psi_i|^2,
\end{equation} represents a non-isolated system in which energy and angular momentum are exchanged
with the environment at some level of fluctuation determined by the ``inverse temperature'', $\beta$, and the ``chemical potential'', $\mu$.  Note that it is possible for the temperature here to be negative but
not the chemical potential for angular momentum.

In their paper, \cite{Lim:2005} show variationally that the 
mean square vortex position (variance) for 2D point vortices,
\begin{equation}
 R^2_{2D} = \langle N^{-1}\sum_i |\Psi_i|^2 \rangle,
\end{equation} where the average is over $P_N$, has a nice formula
\begin{equation}
 R^2_{2D} = \frac{\Omega\beta}{4\mu},
\label{eqn:rsq2d}
\end{equation} where $\Omega=\sum_i\lambda_i$ is the total circulation.  In our case, we define $\lambda_i=1\,\forall i$, so $\Omega=N$ and $R^2_{2D} = N\beta/(4\mu)$.  Moreover, they were
able to show in Monte Carlo simulations of an ensemble of $1000$ point vortices that the
distribution of vortices is almost uniform and axisymmetric, meaning that the probability distribution
of vortices is nearly a perfect cylinder, suggesting that $R^2_{2D}$ is the only value we need to know to determine the entire distribution.

These results provide a tantalizing starting point for our investigation of nearly parallel vortex
filaments where, obviously, at some level of density and fluctuation there exists different
statistical behavior.  Moreover, the discovery of a uniform distribution at low-temperature, we hope,
carries over to nearly parallel vortex filaments.

The low-temperature formula for $R^2_{2D}$, Equation \ref{eqn:rsq2d}, decreases with $\beta$, meaning that the radius
decreases as the ``temperature'', related to the speed of motion of the vortices, increases.  At some
point, of course, $\beta$ is so small that the distribution becomes solely a function of angular
momentum, clearly a normal distribution with variance, $R^2_{2D}=1/(2\mu)$.  The radius does not
decrease to a point, but it \emph{never} increases, certainly not in the low-temperature regime.

The question we address in this paper concerns the statistical distribution of nearly parallel vortex
filaments at low to moderate temperature, \emph{not} high-temperature.  Therefore, our
derivations in Section \ref{sec:meanfield} assume that $\beta$ is large enough that the \emph{fluctuations} in interaction energy and
angular momentum do not play a major role in the statistics, only their mean-values.  The
role of the Monte Carlo comparison in Section \ref{sec:results} is to justify this assumption.  Specifically, we are concerned with the mean square vortex position, $R^2$, which
we define below in Section \ref{sec:meanfield}.

We answer the
questions: (1) do nearly parallel vortex filaments resist and reverse planar collapse as temperature increases (an important question if one would like to contain a column of rotating fluid) and (2) can we see this behavior without violating assumptions of nearly parallel, i.e. do the filaments have to be too curvy before they begin to reverse collapse?  Neither answer is obvious.

\section{Model}
\label{sec:model}
Our model properly derives from the paper of \cite{Lions:2000} and the Gibbs distribution that they introduce.  Although \cite{Klein:1995} derived the equations of motion, we make use of the statistical framework presented in the later paper and also we
rely heavily on their broken-segment model, not only in the Monte Carlo simulation where it is necessary to
discretize the z-axis but also in our mean-field approach.

We define nearly parallel vortex filaments.
\begin{defn}
Nearly parallel vortex filaments are smooth curves that we can represent with a complex parameterization $\Psi_i(\tau)$ where $\Psi_i(\tau)=x_i(\tau) + iy_i(\tau)$ and $\tau\in[0,L)$.  They have a special asymptotic form,  
Given that $L\in O(1)$, if we take any two values of $\tau$, $\tau_0$ and $\tau_1$ such that $\tau_0<\tau_1$, and let
$\Delta\tau=\tau_1-\tau_0$ such that $\Delta\tau\in O(\epsilon)$ where $\epsilon\ll 1$, then
for any filament $i$, the amplitude $|\Psi_i(\tau_1) - \Psi_i(\tau_0)|\in O(\epsilon^2)$.
\end{defn}  In words this means that,
for a small rise of length $\epsilon$ in the filament, the amplitude must be on the order of $\epsilon^2$.  This assumption guarantees a certain degree of straightness in the filament that allows
for the derivation of the quasi-2D equations of motion.  The other asymptotic assumption is of
the vortex core-size, $h$, which has the property $h\ll\epsilon$.  In this model we assume that the
vortices have no cross-section, i.e. the vorticity field, $\omega$, has the form,
\begin{equation}
 \omega = \sum_i\delta(\Psi(\tau) - \Psi_i(\tau)),
\end{equation} where $\Psi(\tau)=x(\tau) + iy(\tau)$ corresponds to the 3D Cartesian position $(x(\tau),y(\tau),\tau)$.

Equilibrium statistical mechanics is traditionally concerned with conserved quantities in a Hamiltonian system.  The Hamiltonian \cite{Klein:1995} derived for nearly parallel vortex filaments has the form,
\begin{equation}
H_N = \alpha\int_0^Ld\tau \sum_{i=1}^{N} \frac{1}{2}\left|\frac{\partial \Psi_i(\tau)}{\partial\tau}
\right|^2 - \int_0^Ld\tau\sum_{i=1}^N\sum_{j>i}^N \log|\Psi_i(\tau) - \Psi_j(\tau)|,
\label{eqn:Ham}
\end{equation} where $\alpha$ is the core-structure constant in units of energy/length.
This Hamiltonian resembles the 2D point vortex Hamiltonian in that the interaction is logarithmic in
the plane.  For two filaments $i$ and $j$, only points in the same plane, the same value of $\tau$, interact.  Two points at different values of $\tau$ only interact if they are both on the same filament.  The first term is a local self-induction term that essentially causes the filaments to wriggle in top-down projection like a particle under Brownian motion.  The additional conserved quantity, angular momentum, is given by,
\begin{equation}
 I_N = \sum_i \int_0^L d\tau |\Psi_i(\tau)|^2.
 \label{eqn:ang}
\end{equation}

\cite{Lions:2000} introduce a broken segment model in Section 2.2, Equation 2.20 of their paper.  For each $i$, $\Psi_i(\tau)$ is piecewise linear with $M$ segments.  Each vertex or ``bead'' is at a multiple of $\delta=L/M$.  Therefore, we define $\psi_i(k) = \Psi_i((k-1)\delta)$, and the curve for
filament $i$ is given by the vector $\Psi_i=(\psi_i(1),\dots,\psi_i(M))$.  With this representation we rewrite
the conserved quantities,
\begin{align}
H_N(M) &= \alpha\sum_{i=1}^{N}\sum_{k=1}^{M} \frac{1}{2}\frac{|\psi_i(k+1) -
\psi_i(k)|^2}{\delta} - \sum_{i=1}^{N}\sum_{j>i}^N\sum_{k=1}^M
\delta\log|\psi_i(k) - \psi_j(k)| ,\\
I_N(M) &= \sum_{i=1}^{N}\sum_{k=1}^M \delta|\psi_i(k)|^2.
\label{eqn:brokenham}
\end{align}

The Gibbs probability measure for a state $s=(\Psi_1,\dots,\Psi_N)$ is then
\begin{equation}
 P(s) = Z^{-1}\exp(-\beta H_N(M) - \mu I_N(M)),
\label{eqn:gibbsProb}
\end{equation} where the partition function has the form,
\begin{equation}
 Z_N(M) = \int_{\C^{MN}} d\Psi_1\cdots d\Psi_N\exp(-\beta H_N(M) - \mu I_N(M)).
\end{equation}  We note that while it is possible to simulate the system with Monte Carlo it is
not possible to solve explicitly for $Z_N$ for any value of $M$.  In their paper, \cite{Lions:2000}
go on to derive a non-linear Schroedinger equation that can give approximate values
for $P(s)$.  Their PDE captures a great deal of the statistics, but they do not provide any explicit formula for the length scale $R$, which is our goal.  We refer the reader to their paper to learn more about their derivation and the model.

\section{Free Energy Theory}
\label{sec:meanfield}
\subsection{Free Energy of Most-Probable Macrostate}
Given a functional for the free energy for a system, $F$, it is possible to solve for the statistics of
the most-probable macrostate by minimizing $F$ with respect to the desired statistic.  In our case the statistic is defined as:

\begin{defn}
The mean square vortex position,
\begin{equation}
 R^2 = \langle L^{-1}N^{-1}\sum_{i=1}^N\int_0^L d\tau |\Psi_i(\tau)|^2 \rangle,
\end{equation} where $\Psi_i(\tau)=x_i(\tau) + iy_i(\tau)$ is a complex number representing the 2D position of filament $i$ at $z=\tau$ and the average $\langle *\rangle$ is with respect to the 3D Gibbs probability measure in $\Psi_i$, 
$P_N=Z_N^{-1}\int d\Psi_1\cdots d\Psi_N \exp(-\beta H_N - \mu I_N)$, where $H_N$ is defined by Equation \ref{eqn:Ham}, $I_N$ by Equation \ref{eqn:ang}.
\label{eqn:rsqDef}
\end{defn}

The difficulty
lies in deriving that functional.  In systems such as ours, with fixed volume ($V=\Re^3$) the Helmholtz description of free energy is appropriate, $F=U - TS$, where $U$ is average energy,
 $T$ is temperature, $S$ is entropy.

In our system we write the free energy as follows:
\begin{equation}
 F_N = \langle H_N \rangle + \frac{\mu}{\beta} \langle I_N \rangle - \frac{1}{\beta}S_N,
\label{eqn:freeEnergy}
\end{equation} where $S=S_N$, $U=\langle H_N \rangle + \frac{\mu}{\beta}\langle I_N \rangle$,
and $T=\frac{1}{\beta}$.

While many statistical mechanics approaches are devoted to finding a functional for $S_N$, since $H_N$ and $I_N$ are
usually known, there is another way to develop the free energy functional from
the partition function $Z_N$.  The function $Z_N$ can be considered a sum over microstates, ${s_i}$, or it can be an average over a set of macrostates, ${\sigma_j}$.  Equation \ref{eqn:partfn} is a sum over microstates.  Alternatively, the partition function is given by the formula
\begin{equation}
 Z_N = \sum_j \exp(-\beta H_N[\sigma_j] - \mu I_N[\sigma_j])P[\sigma_j],
\end{equation} where $H_N$ is the energy, $I_N$ is the angular momentum, and $P$ is the probability for macrostate $\sigma_j$.  

Since $S_N[\sigma_j]=\log P[\sigma_j]$, using the formula \ref{eqn:freeEnergy}, we can say,
\begin{align}
 Z_N &= \sum_j \exp(-\beta H_N[\sigma_j] - \mu I_N[\sigma_j] + S[\sigma_j])\nonumber\\
 &= \sum_j \exp(-\beta F_N[\sigma_j]).
\end{align}
Because of conservation laws, we assume
in physics that the most-probable macrostate or energy-state, $j=m$, has a probability so much larger than the probabilites of
all other macrostates that sum contributions from other macrostates can be neglected and
\begin{equation}
 Z_N = \exp(-\beta F_N[\sigma_m]).
\end{equation}
Therefore, we have an equation for the free energy,
\begin{equation}
 F_N = -\frac{1}{\beta}\log Z_N.
\label{eqn:freeEnergy2}
\end{equation} Even though this free energy is only the free energy of the most-probable
macrostate, it can be considered the system's free energy.

\subsection{Mean Field Assumption}
As mentioned in Section \ref{sec:model}, we cannot solve for $Z_N$.  We need to approximate it
to derive the free energy functional.  One way to do this is with a mean-field assumption.  Because
the logarithmic interaction in the Hamiltonian prevents us from evaluating the integral over
microstates, Equation \ref{eqn:partfn}, we need to remove the interdependence of filaments upon one-another so that we can separate the integrals over different filaments.  In their paper
this is the approach of \cite{Lions:2000}, in that they develop a model of a filament that interacts
with a probability distribution, or field, rather than other filaments.  Our approach is to take a
cue from the physics of Newtonian gravity and assume that the filament interacts with the other filaments in the same way its center of mass would interact with an imaginary, ``center of mass'' filament.  We motivate this assumption from the Monte Carlo results on point vortices of \cite{Lim:2005} that showed a ``flat-top'', cylindrical probability distribution for vortices in the plane.  A nearly uniform, symmetric distribution is crucial to the center-of-mass assumption.

Given a point on filament $i$, $\psi_i(\tau)$, if the filament's center-of-mass interacts with the center of mass of the other filaments, the interaction potential, $V_{i}$, simplifies:
\begin{align*}
 V_{i}(\tau) &= \sum_{j} -\hf\log|\psi_i(\tau) - \psi_j(\tau)|\\
 &= -(N-1)\hf\log|\psi_i(\tau)| = -N/4\log|\psi_i(\tau)|^2\\
&= -N/4\log \left(L^{-1}\int_0^L d\tau |\psi_i(\tau)|^2\right)\\
&= -N/4\log \frac{I_N}{L},
\end{align*} where we can say for large $N$ that $N-1\sim N$.  This result makes the interaction
a function of the angular momentum.

The center-of-mass assumption liberates us in the evaluation of the partition function, because we need to consider \emph{neither} the configurations of other filaments \emph{nor} their density distribution, whereas Lions and Majda do take the latter into account.  All we need consider is the angular
momentum.

\subsection{Spherical Constraint}
Given the mean-field, center-of-mass assumption, we have a new Hamiltonian system governing the
behavior of $N$ independent, $M$-segment filaments.  The Hamiltonian reads

\begin{align}
H_N^{cm}(M) &= \alpha\sum_{i=1}^{N}\sum_{k=1}^{M} \frac{1}{2}\frac{|\psi_i(k+1) -
\psi_i(k)|^2}{\delta} - N/4\sum_{i=1}^{N}\sum_{k=1}^M
\delta\log \frac{I_N}{L},
\end{align} where $H_N^{cm}(M)$ only differs from $H_N(M)$ (Equation \ref{eqn:brokenham}) in the second term, and the new partition function reads
\begin{align}
 Z_N^{cm}(M) &= \int d\Psi_1\cdots d\Psi_N \exp(-\beta H_N^{cm}(M) - \mu I_N^{cm}(M))\nonumber\\
&= \left\{\int d\Psi_1 \exp(-\beta H_1^{cm}(M) - \mu I_1^{cm}(M))\right\}^N,
\end{align} because all the integrals are independent and equal. 

Returning to our derivation for the length scale, $R$, we take the simplest course possible to give a reasonable formula and impose a constraint on the filament, (dropping
the $1$ subscript):
\begin{align}
 L^{-1}\int_0^L d\tau |\Psi(\tau)|^2 = R^2 = \sum_{k=1}^M M^{-1}|\psi(k)|^2,
\end{align} which imposes an average distance from the origin over the length of the filament.  We include this constraint as a micro-canonical or exact constraint into the partition
function,
\begin{equation}
 Z_N^{cm,sc}(M) = \left\{\int d\Psi e^{\left(-\beta H_1^{cm}(M) - \mu I_1^{cm}(M)\right)}
\delta\left(R^2 - \sum_{k=1}^M M^{-1}|\psi(k)|^2\right)\right\}^N.
\label{eqn:partfn_cm_sc}
\end{equation}  This constraint eliminates fluctuations in interaction energy and angular momentum,
leaving only fluctuations in local self-induction energy.  This elimination provides the best
opportunity for comparison with the 2D length scale work of \cite{Lim:2005} that also eliminates
these two types of fluctuation.

Equation \ref{eqn:partfn_cm_sc} is a multi-dimensional Gaussian integral with a spherical constraint---so called because it forces the vector $\Psi$ to stay on the $M-1$-sphere of radius $\sqrt{M}R$.  To evaluate it we turn to the spherical model of \cite{Berlin:1952}.

The spherical model comprises a number of steps for evaluating integrals of the form of
Equation \ref{eqn:partfn_cm_sc}, beginning by putting the Dirac-delta function into integral form and ending with a steepest descent evaluation of the integral over $\Psi$.

In integral (Fourier) form the Dirac-delta reads:
\begin{equation}
 \delta\left(MR^2 - \sum_{k=1}^M |\psi(k)|^2\right) = 
\int_{-\infty}^{\infty}\frac{d\sigma}{2\pi}\exp\left[-i\sigma\left(MR^2 - \sum_{k=1}^M |\psi(k)|^2\right)\right],
\label{eqn:sphere}
\end{equation} and allows us to combine the function in the exponent in Equation \ref{eqn:partfn_cm_sc}
with the exponent of the spherical constraint.  (Whereas before $\sigma$ was used as a symbol for
macrostate, here it is an integration variable.)  

\subsection{Free Energy Derivation for $R$}
Now we can determine the free energy functional, which ought to be minimal under the constraints of the system, via steepest descent.  Determining the free energy as a function of the particular statistic, $R$, is more useful than having an equation for the partition function itself because we can determine $R$ by minimizing the free energy w.r.t. it and then solving for $R$.  
\begin{thm}
Given the partition function defined in 
Equation \ref{eqn:partfn_cm_sc} and the free energy in Equation \ref{eqn:freeEnergy}, as $M\rightarrow\infty$, the value for $R^2$ giving minimal free energy is,
 \begin{equation}
 R^2 = \frac{\beta^2\alpha N + \sqrt{\beta^4\alpha^2N^2 + 32\alpha\beta\mu}}{8\alpha\beta\mu}.
\end{equation}
\end{thm}

\begin{proof}
The steepest descent method is an excellent way to obtain values for the
minimum free energy.  In employing steepest descent, we need to start with an integral of the form, $\int dx e^{-MF[x]}$.  Then if $F[x_0]< F[x]\forall x\neq x_0$, i.e. its minimum value is at $x_0$,
\[
\lim_{M\rightarrow\infty} M^{-1}\log\left(\int dx e^{-MF[x]}\right) = F[x_0].
\]  This works because, as $M$ increases, the distribution that $e^{-MF[x]}$ represents
becomes narrower and focuses on $x_0$ until the distribution has zero value at all other $x$.

To take this approach, we put the partition function in the appropriate form:  If
\begin{equation}
 Z_N^{cm,sc}(M)= \left\{\int d\Psi\int_{-\infty}^{\infty}\frac{d\sigma}{2\pi} e^{-MF[i\sigma]}\right\}^N,
\label{eqn:partfn_new}
\end{equation} then
\begin{align}
 F[i\sigma] = &\alpha\beta M^{-1}\sum_{k=1}^{M} \frac{1}{2}\frac{|\psi(k+1) -
\psi(k)|^2}{\delta} - N\beta LM^{-1}/4\log R^2 \nonumber\\&+ \mu LM^{-1}R^2 - i\sigma\left(R^2 - \sum_{k=1}^M M^{-1}|\psi(k)|^2\right),
\end{align} is the non-dimensional free energy, where we have already applied the Dirac delta function
to the interaction energy and the angular momentum.  

We can pull the constant terms out of the integral over $\Psi$.  Because $F$ is positive definite, under Fubini's theorem we may switch
the integrals to obtain
\begin{equation}
 Z_N^{cm,sc}(M)= \left\{e^{N\beta L/4\log R^2 - \mu LR^2}\int_{-\infty}^{\infty}\frac{d\sigma}{2\pi}
\int d\Psi e^{-F'[i\sigma]}\right\}^N,
\label{eqn:partfn_new2}
\end{equation} where
\begin{equation}
 F'[i\sigma] = \alpha\beta \sum_{k=1}^{M} \frac{1}{2}\frac{|\psi(k+1) -
\psi(k)|^2}{\delta} + i\sigma\left(\sum_{k=1}^M |\psi(k)|^2 - MR^2\right),
\end{equation} is the free energy still dependent on $\Psi$.

The interior integral needs evaluation.  Let $s=i\sigma$ and
\begin{equation}
 Z'(M) = \int_{-i\infty}^{i\infty}\frac{ds}{2\pi}\int d\Psi e^{-F'[s]},
\label{eqn:interiorZ}
\end{equation} be that interior.  Let us put $F'$ in matrix form:
\begin{equation}
 F'[s] = -sMR^2 + K\Psi^\dagger A \Psi + s\Psi^\dagger\Psi,
\end{equation} where
$K=\alpha\beta/\delta$, and the $M\times M$ matrix $A$ has the form
\begin{align*}
A_{i,i} &= 1,\\
A_{i,i+1} = A_{i+1,i} = A_{1,M} = A_{M,1} &= -\hf,\\
A_{i,j} &= 0\quad other\quad i,j.
\end{align*}

The integral in Equation \ref{eqn:interiorZ} is Gaussian.  We can evaluate it, knowing
the eigenvalues of the matrix $A$.  These eigenvalues have the form
$\lambda_i=1 - \cos(2\pi (i-1)/M)$, (not related to the previous use of $\lambda_i$ as strength of vorticity)(\cite{Berlin:1952},\cite{Lions:2000}) and so
\begin{align}
 Z'(M) &= \int_{-i\infty}^{i\infty}\frac{ds}{2\pi}e^{sMR^2}\pi^M\prod_i \frac{1}{s + K\lambda_i}.
\end{align}

We need to put $Z'(M)$ back into the correct form for steepest descent.  Following the example of Berlin and Kac, let $s=K(\eta-1)$ then
\begin{equation}
 Z'(M) = \int_{-i\infty}^{i\infty}\frac{d\eta}{2\pi}K\pi^M(\eta - 1)^{-1}e^{-M\log(K)}e^{Mf[\eta]},
\end{equation} where
\begin{equation}
f[\eta] = KR^2(\eta-1) - M^{-1}\sum_{i=2}^M \log(\eta - \cos(2\pi (i-1)/M)),
\end{equation} and $\eta\geq 1$.

We leave the $i=1$ term out of the sum in $f[\eta]$ so that we can evaluate $f$ further by taking the limit on the second term,
\[
\lim_{M\rightarrow\infty} M^{-1}\sum_{i=2}^M \log(\eta - \cos(2\pi (i-1)/M)) = 
\frac{1}{2\pi}\int_0^{2\pi} d\omega \log(\eta - cos(\omega)),
\] which gives
\begin{align}
 f[\eta] &= KR^2(\eta-1) - \frac{1}{2\pi}\int_0^{2\pi} d\omega \log(\eta - cos(\omega))\nonumber\\
&= KR^2(\eta-1) - \log(\eta + (\eta^2 - 1)^\hf).
\end{align}

To apply steepest descent, we determine the saddle point $\eta=\eta_0$ where $f[\eta]$ has its minimum value, $f[\eta_0]$.  Taking the derivative and setting it equal to zero,
\begin{equation}
 \frac{\partial f}{\partial \eta} = KR^2 - \frac{1}{\sqrt{\eta^2 - 1}} = 0,
\end{equation} implies
\begin{equation}
\eta_0 = \sqrt{\frac{1}{(KR^2)^2} + 1}.
\end{equation}

Having evaluated $f$, we can give an equation for the original free energy,
\begin{align}
 F[\eta_0] &= -N\beta L/4\log R^2 + \mu LR^2 + M\log(K) - Mf[\eta_0],
\end{align} and evaluate it as $M\rightarrow\infty$.

The term in the limit $M\log(K)$, does not depend on $R^2$, and it is an unnecessary component representing the entropy of the
broken filaments in the non-interacting case, and we drop it.  Now we fill
in the expressions for $\eta_0$ and $K$ as defined above:
\begin{align}
 \lim_{M\rightarrow\infty} Mf[\eta_0] &= \lim_{M\rightarrow\infty} KR^2(\eta_0-1) - M\log(\eta_0 + (\eta_0^2 - 1)^\hf)\nonumber\\
&= \lim_{M\rightarrow\infty} MKR^2\left(\sqrt{\frac{1}{(KR^2)^2} + 1} - 1\right) - M\log\left(\sqrt{\frac{1}{(KR^2)^2} + 1} + \frac{1}{KR^2}\right)\nonumber\\
&= \lim_{M\rightarrow\infty} M\frac{\alpha\beta M}{L}R^2\left(\sqrt{\frac{1}{(\frac{\alpha\beta M}{L}R^2)^2} + 1} - 1\right) - M\log\left(\sqrt{\frac{1}{(\frac{\alpha\beta M}{L}R^2)^2} + 1} + \frac{1}{\frac{\alpha\beta M}{L}R^2}\right)
\end{align} which, because it is an energy for a filament, ought to be finite.  The first term is the energy of the filament, $E_{fil}$, and the second, the entropy, $S_{fil}$, and each by itself is finite, so we take each limit separately.

The energy limit is simple to calculate using L'H\^opital's rule:
\begin{align}
E_{fil} &= \lim_{M\rightarrow\infty} M\frac{\alpha\beta M}{L}R^2\left(\sqrt{\frac{L^2}{(\alpha\beta MR^2)^2} + 1} - 1\right) \nonumber\\
&= \lim_{\delta\rightarrow 0} L\frac{\alpha\beta}{\delta^2}R^2\left(\sqrt{\frac{\delta^2}{(\alpha\beta R^2)^2} + 1} - 1\right)\nonumber\\
&= \lim_{\delta\rightarrow 0} L\frac{\alpha\beta}{2\delta}R^2\left( \left( \frac{\delta^2}{(\alpha\beta R^2)^2} + 1\right)^{-\hf}\frac{\delta}{(\alpha\beta R^2)^2}\right) \nonumber\\
&= \frac{L}{2\alpha\beta R^2}
\end{align}

The entropy limit is equally simple:
\begin{align}
 S_{fil} &= \lim_{M\rightarrow\infty} M\log\left(\sqrt{\frac{1}{(\frac{\alpha\beta M}{L}R^2)^2} + 1} + \frac{1}{\frac{\alpha\beta M}{L}R^2}\right)\nonumber\\
&= \lim_{\delta\rightarrow 0} \frac{L}{\delta}\log\left(\sqrt{\frac{\delta^2}{(\alpha\beta R^2)^2} + 1} + \frac{\delta}{\alpha\beta R^2}\right)\nonumber\\
&=\lim_{\delta\rightarrow 0} L\left[\sqrt{\frac{\delta^2}{(\alpha\beta R^2)^2} + 1} + \frac{\delta}{\alpha\beta R^2}\right]^{-1}\left[\left(\frac{\delta}{(\alpha\beta R^2)^2} + 1\right)^{-\hf}\frac{\delta}{(\alpha\beta R^2)^2} + \frac{1}{\alpha\beta R^2}\right]\nonumber\\
&=\frac{L}{\alpha\beta R^2}
\end{align}

These two results imply that
\[
 \lim_{M\rightarrow\infty} Mf[\eta_0] = -\frac{L}{2\alpha\beta R^2}
\] and
\begin{equation}
 F[\eta_0] = L\mu R^2 - N\beta L/4\log R^2 + \frac{L}{2\alpha\beta R^2}.
\end{equation}

We minimize with respect to $R^2$,
\begin{align}
 \frac{\partial F}{\partial R^2} = L\mu - \frac{N\beta L}{4 R^2} - \frac{L}{2\alpha\beta R^4} = 0,
\end{align} and solve for $R^2$,
\begin{align}
 R^2 &= \frac{N\beta/4 \pm \sqrt{(N\beta/4)^2 + 4\mu\frac{1}{2\alpha\beta}}}{2\mu}\nonumber\\
&= \frac{\beta^2\alpha N + \sqrt{\beta^4\alpha^2N^2 + 32\alpha\beta\mu}}{8\alpha\beta\mu},
\end{align} where we take the ``plus'' solution as giving physical results.
\end{proof}
We note that the relationship between $F[\eta_0]$ and our previous $F_N$ in Equation \ref{eqn:freeEnergy} is $F[\eta_0]=\beta F_N$, meaning that $F[\eta_0]$ is non-dimensional while
$F_N$ has units of energy.

The resulting expression for the square length scale $R^2$ is useful for comparison with the
length scale result of \cite{Lim:2005} because, if we take the limit
\begin{equation}
 \lim_{\alpha\rightarrow\infty} \frac{\beta^2\alpha N + \sqrt{\beta^4\alpha^2N^2 + 32\alpha\beta\mu}}{8\alpha\beta\mu} = \frac{N\beta}{4\mu},
\end{equation} we get back the 2D point vortex result for the length scale, which shows that
our formula and the \cite{Lim:2005} formula agree for perfectly straight filaments.

However, for finite $\alpha$ the two formulae show a signficant difference.  The 2D formula is
linear in $\beta$, our formula is non-linear.  In fact, for decreasing $\beta$, the sign of the slope of our
formula changes at $\beta=\beta_0$, where
\begin{equation}
 \beta^3_0 = \frac{4\mu}{\alpha N^2}.
\end{equation}  That the system collapses and then starts to expand as ``temperature'', $1/\beta$, increases indicates a significant departure from the strictly-2D where the system size only collapses.  In Section \ref{sec:results},
we show that that Monte Carlo confirms this result and that the straightness assumptions of the model
hold through much of the expansion phase.

\section{Monte Carlo}
\label{sec:montecarlo}
Path Integral Monte Carlo methods emerged from the path integral formulation invented by Dirac 
that Richard Feynman later expanded (\cite{Zee:2003}), in which particles are conceived to follow
all paths through space.  One of Feynman's great contributions to
the quantum many-body problem was the mapping of path integrals onto a classical system of interacting
``polymers'' (\cite{Feynman:1948}).  D. M. Ceperley used Feynman's convenient piecewise linear formulation to develop his PIMC method which 
he successfully applied to He-4, generating the well-known lambda transition for the first time
in a microscopic particle simulation (\cite{Ceperley:1995}). Because it describes a system
of interacting polymers, PIMC applies to classical systems that have a ``polymer''-type
description like nearly parallel vortex filaments.  

PIMC has several advantages.  It is a
\emph{continuum} Monte Carlo algorithm, relying on no spatial lattice.  Only time (length in the z-direction
in the case of vortex filaments) is discretized, and the algorithm makes no assumptions about types of phase transitions or trial wavefunctions.

\begin{figure}
\begin{center}
\includegraphics[width = 0.5\textwidth]{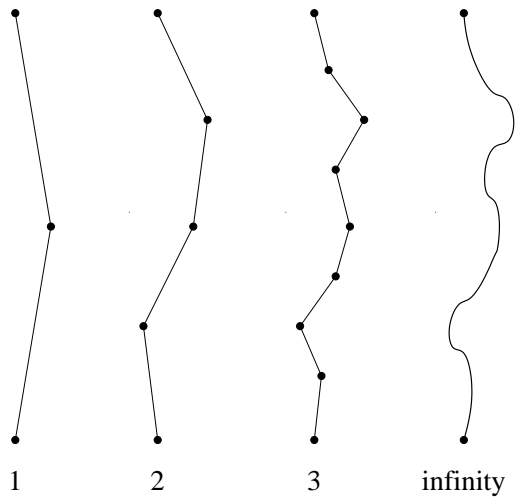}
\end{center}
\caption{The bisection algorithm works by bisecting the filament to sample point positions.  First
the center point is selected, then the two points half-way from the center to the end-points, then four more points, eight, and so on
until some maximum number of points are sampled.  These are snapshots of one filament at different steps in the sampling process.}
\label{fig:bisection}
\end{figure}

The Monte Carlo simulation begins with a random distribution of filament end-points in a square of side
10, and there are two possible moves that the algorithm chooses at random:  
\begin{enumerate}
 \item Moves a filament's end-points, $\psi_i(1)$ and $\psi_i(M+1)$.  The index $i$ is chosen at random, and the filament $i$'s end-points moved a uniform random distance.
 \item Keeps end-points stationary and, following the bisection method of Ceperley (Figure \ref{fig:bisection}), grows
a new internal configuration for a randomly chosen filament
\end{enumerate} (\cite{Ceperley:1995}).  

In each case, the energy of the new state, $s'$, is calculated and retained with probability
\begin{equation}
 A(s\rightarrow s') = \min\left\{1,\exp\left(-\beta [H_N^{s'}(M)-H_N^s(M)] - \mu [I_N^{s'}(M)-I_N^s(M)]\right)\right\},
\end{equation}  where $s$ is the previous state.  This effectively samples states from the Gibbs
probability distribution in Equation \ref{eqn:gibbsProb}.

Our stopping criteria is graphical in that we ensure that the cumulative arithmetic mean of the energy,
\begin{equation}
E_{cum}^k = k^{-1}\sum_{i=1}^k H_N(s_i) + \frac{\mu}{\beta}I_N(s_i),
\end{equation} where $s_i$ refers to the state resulting from the $i$th move and $k$ is
the current move index,
 settles to
a constant.  The energy is almost guaranteed to settle in the case of the Gibbs' measure because of the tendency for the system to select a particular energystate (mean energy) and remain close to that state.  Typically, we run for 1 million moves (accepted plus rejected) or 50,000 sweeps for 20 vortices.  Afterwards, we
collect data from about 200,000 moves (1000 sweeps) to generate statistical information.

\section{Results}
\label{sec:results}
\subsection{Comparison}
We simulated a collection of $N=20$ vortices each with a piecewise linear representation with $M=1024$ segments and ran the system to equilibration, determined by the settling of the mean and variance of the total
energy.  We ran the system for 20 logarithmically spaced values of $\beta$ between $0.001$ and $1$ plus two points, $10$ and $100$.  We set $\alpha=10^7$ (enforcing straightness), $\mu=2000$, and $L=10$.  Decreasing $\beta$ simulates an increase in temperature, $1/\beta$.  We calculate two arithmetic averages: the mean square vortex position,
\begin{equation}
 R^2_{MC} = (MN)^{-1}\sum_{i=1}^N\sum_{k=1}^M|\psi_i(k)|^2,
\label{eqn:rsqMC}
\end{equation} and the mean
square amplitude per segment,
\begin{equation}
 a^2 = (MN)^{-1}\sum_{i=1}^N\sum_{k=1}^M|\psi_i(k)-\psi_i(k+1)|^2, 
\label{eqn:aps}
\end{equation} where $\psi_i(M+1)=\psi_i(1)$.

\begin{figure}
\begin{center}
\includegraphics[width = \textwidth]{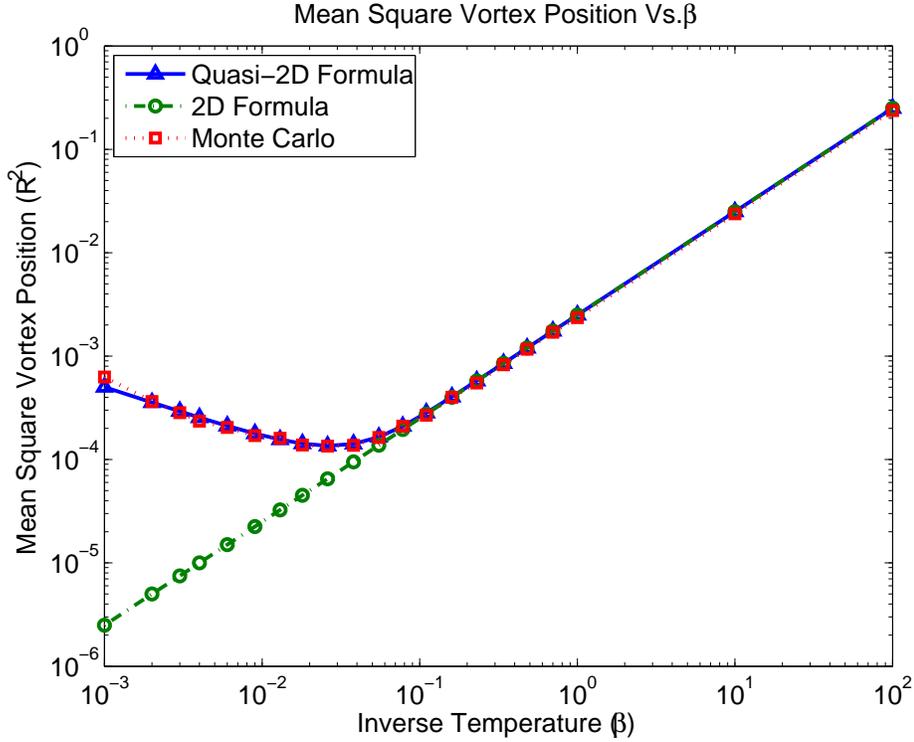}
\end{center}
\caption{The mean square vortex position, defined in Equation \ref{eqn:rsqMC}, compared with
Equations \ref{eqn:rsq3d} and \ref{eqn:rsq2d} shows how 3-D effects come into play around
$\beta=0.16$.  That the 2D formula continues to decrease while the Monte Carlo and the quasi-2D
formula curve upwards with decreasing $\beta$ suggests that the internal variations of the
vortex lines have a significant effect on the probability distribution of vortices.}
\label{fig:mf2mcComp}
\end{figure}

Measures of the Monte Carlo $R^2_{MC}$, Equation \ref{eqn:rsqMC}, correspond well to the 3D $R^2$, Equation \ref{eqn:rsq3d}, in Figure 
\ref{fig:mf2mcComp} whereas, the strictly-2D $R^2_{2D}$, Equation \ref{eqn:rsq2d}, continues to decline when the others curve
up with decreasing $\beta$ values, suggesting that the 3-D effects are not only real in the Monte Carlo
but that the mean-field is a good approximation with these parameters.  In Section \ref{sec:discussion}
we discuss what this expansion really means.

\subsection{Straightness Holds}
\label{sec:straightness}

\begin{figure}
\begin{center}
\includegraphics[width = 0.6\textwidth]{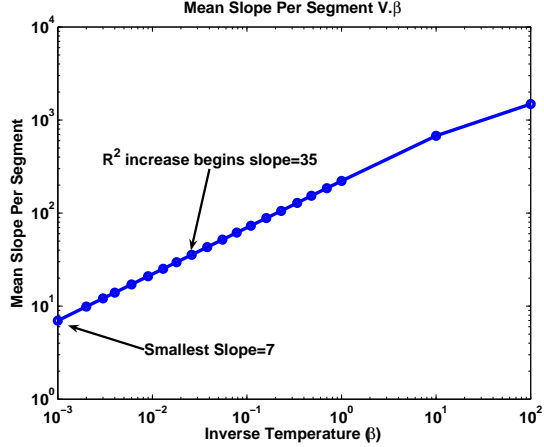}
\end{center}
\caption{This figure shows the mean slope per segment, $\delta/a$, where $\delta\sim 0.0098$ and $a$ is given by Equation \ref{eqn:aps}, and that
straightness holds for all $\beta$ values.  The point at which $R^2$ begins to increase with decreasing $\beta$ (Figure \ref{fig:mf2mcComp}) has a mean slope of $35$, and, even at the smallest $\beta=10^{-3}$, the
segments have an average angle of $82^{\circ}$ with respect to the complex plane.}
\label{fig:check1}
\end{figure}

In order to be considered straight enough, we need
\begin{equation}
a \ll \frac{L}{M} = \frac{10}{1024} \sim 0.0098.
\end{equation}  Straightness holds for all $\beta$ values, shown in Figure \ref{fig:check1}.  These conditions hold on average.  We ignore extreme low-probability cases as not contributing significantly to the statistics.  These straightness constraints, coupled with filaments having no attractive interactions, mean hairpins (kinks or, in quantum terminology, instantons) do not occur.

Aside from straightness constraints, the reader might question whether allowing vortex filaments
to entangle violates the model's assumptions.  While this is a valid concern, it is not an assumption of the model.  Since this fluid is almost-everywhere invicid, it contains asymptotically small regions of
non-zero viscosity, and, while a totally invicid fluid cannot allow vortices to change topology
(cross over each other) from state to state, an almost-everywhere invicid fluid allows vortex
reconnections and cross-overs to occur due to microscopic viscous effects.  Therefore, in our
simulations vortices are allowed to cross one-another.  We are not
claiming to model vortex reconnection, which is a mysterious process, but only the before and after
effects of it.

Concerning the question of how vortices can cross one-another and still remain nearly parallel, we
point to the extremely high-density (tiny value of $R^2$) which allows even the straightest filaments
to entangle.

\section{Discussion}
\label{sec:discussion}
We have shown that with increasing $\beta$ there is a ``transition'' from decreasing $R^2$, the
2D behavior, to increasing $R^2$, the 3D behavior.  We put this word ``transition'' in quotes because
at present we have no proof for or against this being a phase transition.  The free energy function
that we derive is smooth for all positive $\beta$.  However, this is not proof against there being
a phase transition in the original system.  
At this
point it is a subject for future research.

The decreasing-$\beta$, $R$-expansion suggests that, by adding degrees of freedom to the 2D model to make it a
quasi-2D model, we add a mechanism for the vortices to resist confinement through entropic effects.
As $\beta$ decreases past the ``transition'', the system's $R^2$ goes from being the result of interaction-versus-angular momentum competition to an entropy-versus-angular momentum competition.
In the 2D system this 3rd dimension entropy is not there.  Although there is another kind of entropy in the 2D
model that will slow the compression as $\beta\rightarrow 0$ to a constant value, it is not enough
to cause an increase in the system size.  In the 3D system, the degrees of freedom are exponentially greater, which causes the expansion seen in Figure \ref{fig:mf2mcComp}.

\section{Conclusion}
\label{sec:conc}
Statistical mechanics provides a way to model transitions in decaying turbulence when there is a separation of time scales.  We conclude from our results that the transfer of the model size from an interaction-angular momentum competition to an entropy-angular momentum competition is a kind of transition to turbulence purely due to 3D effects.  Whether it is a phase transition is unknown, but it shows that 3D effects do become significant in determining the length scale of the turbulent system, a useful result for experiments in tight confinement of rotating nearly-invicid fluids.

The key role of angular momentum in the derivation of the expression for $R^2$ and the subsequent
Monte Carlo validation cannot be over-emphasised.  At positive $\beta$ and $\alpha$, quasi-2D
vortices of the same sign effectively repulse one-another.  Therefore, in the unbounded plane, they
would fly off to infinity without the angular momentum constraint unless hemmed in by an infinite
expanse of vortices, the case of periodic boundaries, or walls.  As we mentioned in the introduction,
walls and periodic boundaries enforce an artificial length scale on the vortices, while angular
momentum constraints allow that length scale to be found naturally.  

In experiments and simulations of experiments, it is often reasonable to model walls present
in the experimental setup.  However, in oceans and atmospheres and in stars, there are no walls, and,
at scales where Coriolis effects are small, it does not make sense to use a spherical domain unless
the simulation is of extremely high resolution.  The proper regime for applications of unbounded plane, conserved angular
momentum simulations is in small Rossby number regions where small-scale rotation occurs such
as \cite{Julien:1996} have modeled in their astrophysical simulations and found arrays of
like-sign nearly parallel vortex filaments.  To this niche we have contributed a previously
unseen entropic transfer to turbulence in rotating, almost-everywhere ideal fluids.

\centering
\large
{\bf Acknowledgments}
\flushleft
\normalsize
This work is supported by ARO grant W911NF-05-1-0001 and DOE grant 
DE-FG02-04ER25616. 
\pagebreak
\bibliography{pimcposter}

\end{document}